\documentclass[format=acmsmall, review=false, screen=true, printfolios=true]{acmart}

\renewcommand\footnotetextcopyrightpermission[1]{}

\usepackage{booktabs}
\usepackage{pgfplots}
\pgfplotsset{compat=newest}
\usepackage{xcolor}
\usepackage{siunitx}
\usepackage{multirow}
\usepackage{xspace}
\usepackage{balance}
\usepackage[font=rm]{caption}
\usepackage{verbatim}
\usepackage{graphicx}
\usepackage{tabularx}
\usepackage{enumitem}
\usepackage{shortvrb}
\usepackage{tikz}
\usepackage{subfig}
\usepackage[boldmath]{numprint}
\usepackage{listings}

\usepackage[noend]{algpseudocode}
\usepackage[linesnumbered
			,vlined]{algorithm2e}
\let\oldnl\nl
\newcommand{\nonl}{\renewcommand{\nl}{\let\nl\oldnl}}

\makeatletter
\newcommand{\removelatexerror}{\let\@latex@error\@gobble}
\makeatother

\setlist[itemize,1]{leftmargin=5mm,itemsep=0mm}
\setlist[enumerate,1]{leftmargin=8mm,itemsep=0mm}

\definecolor{keywords}{HTML}{44548A}
\definecolor{strings}{HTML}{00999A}
\definecolor{comments}{HTML}{990000}

\definecolor{DarkGray}{gray}{.45}

\newcommand{\return}{\vspace{0.3cm}}
\newcommand{\parag}[1]{\return\noindent\textbf{\textsf{#1.}}}

\newcommand{\aftertabspace}{\vspace{0.15cm}}

\newcommand{\mytablescale}{0.95} 

\SetKwIF{If}{ElseIf}{Else}{{\textsf{if}}}{}{\textsf{else if}}{{\textsf{else}}}{}
\SetKwFor{For}{\textsf{for}}{}{}
\SetKwFor{While}{\textsf{while}}{}{}
\newcommand{\code}[1]{\texttt{\textbf{#1}}}
\newcommand{\func}[1]{\texttt{\textbf{#1}}}

\usepackage{xcolor, soul}

\newcommand{\perf}[2]{{$#1\times$ $\div$ $#2\times$}}

\usepackage{pbox}
\usepackage{makecell}
\usepackage{tikz}
\usetikzlibrary{calc}

\newcommand{\gov}{\textsf{Gov2}}
\newcommand{\clue}{\textsf{CW09}}
\newcommand{\cc}{\textsf{CCNews}}

\newcommand{\RNum}[1]{\uppercase\expandafter{\romannumeral #1\relax}}

\newcommand{\seq}{\mathcal{S}}

\newcommand{\access}{\textup{\textsf{access}}}
\newcommand{\Nextgeq}{\textup{\textsf{nextGEQ}}}

\newcommand{\method}[1]{{\sf{#1}}}

\newcommand{\var}[1]{\mbox{\emph{#1}}}

\newcommand{\interp}{{\method{BIC}}}
\newcommand{\vb}{{\method{V}}}
\newcommand{\opt}{{\method{PEF}}}
\newcommand{\uniform}{{\method{EF}}}
\newcommand{\roar}{{\method{R2}}}
\newcommand{\roaropt}{{\method{R3}}}
\newcommand{\slice}{{\method{S}}}

\newcommand{\pef}{{\method{PEF}}}

\begin{document}
\title{On Slicing Sorted Integer Sequences}

\author{Giulio Ermanno Pibiri}
\orcid{0000-0003-0724-7092}
\affiliation{%
  \institution{ISTI-CNR}
  \city{Pisa}
  \state{Italy} 
}

\email{giulio.ermanno.pibiri@isti.cnr.it}

\begin{abstract}

\medskip

Representing sorted integer sequences in small space
is a central problem for large-scale retrieval
systems such as Web search engines.
Efficient query resolution, e.g., intersection or
random access,
is achieved by carefully \emph{partitioning} the sequences.

In this work we describe and compare
two different partitioning paradigms:
partitioning \emph{by cardinality}
and partitioning \emph{by universe}.
Although the ideas behind such paradigms have been known in the
coding and algorithmic community since many years,
inverted index compression has extensively adopted
the former paradigm, whereas the latter has received only
little attention.
As a result, an experimental comparison between these two
is missing for the setting of inverted index compression.

We also propose and implement a solution that
recursively slices the universe of representation of
a sequence to achieve compact storage and attain to
fast query execution.
Albeit larger than some state-of-the-art representations,
this \emph{slicing} approach substantially improves
the performance of list intersections and unions while
operating in compressed space,
thus offering an excellent space/time trade-off
for the problem.

\end{abstract}

\maketitle

\section{Introduction}\label{sec:introduction}

Large-scale
retrieval systems employ a simple, yet ingenious,
data structure to support text search
-- the
\emph{inverted index}~\cite{ZobelM06,Raghavan-book,Cambazoglu:2015,mg}.
In its simplest incarnation, the inverted index
is a collection of sorted integer sequences,
called inverted lists.
For each distinct term appearing in the textual
collection, the corresponding inverted list
represents the list of the identifiers of the
documents where the term appears.
Then, resolving a user query such as,
for example, \emph{``return all documents where terms
$t_1$ and $t_2$ appear''}
reduces to the problem of \emph{intersecting}
the inverted lists of $t_1$ and $t_2$.
Other query operators are possible and
several pruning techniques have been developed~\cite{Wand,BlockMaxWand}
for the case of \emph{ranked} retrieval, i.e.,
when the returned
documents have to be ranked according to
a scoring function~\cite{robertson1976relevance}.
~\citet*{ZobelM06} provide general background
on inverted indexes.

\return
Literature on the representation of integers and
integer sequences is vast.
Many solutions
are known, each of them exposing a different space/time
trade-off, including:
Elias' gamma and delta~\cite{elias75},
Golomb~\cite{Golomb66},
Elias-Fano~\cite{Fano71,Elias74,2013:vigna},
partitioned Elias-Fano~\cite{2014:ottaviano.venturini},
clustered Elias-Fano~\cite{2017:pibiri.venturini},
Interpolative~\cite{1996:moffat.stuiver,2000:moffat.stuiver},
PForDelta~\cite{zukowski06super,2009:yan.ding.ea,2013:lemire.boytsov},
Simple~\cite{AnhM05,zlt08www,AnhM10},
Variable-Byte~\cite{thiel1972program,scholer2002compression,2009:dean,2011:stepanov.gangolli.ea,2015:plaisance.kurz.ea,2018:lemire.kurz.ea,TKDE19},
QMX~\cite{2014:trotman},
ANS-based~\cite{ANS1,ANS2},
DINT~\cite{DINT}.
We point the reader to the surveys by~\citet*{ZobelM06},
by~\citet*{2008:moffat} and by~\citet*{EBDT2018}
for a review of many techniques.

\return
More precisely, the problem we take into account is the
one of introducing a compressed representation for a
sorted integer sequence $\seq(n,u)$ of size $n$ whose values
are drawn from a universe $u \geq \seq[n-1]$,
here assumed to be strictly increasing, i.e.,
$\seq[i] > \seq[i-1]$ for $0<i<n$, so that the following operations have to be
supported efficiently.

\begin{itemize}

\item $\seq$.\textsf{decode}(\var{output}):
decodes $\seq$ sequentially to the \var{output} buffer
of 32-bit integers;

\item \textsf{AND}/\textsf{OR}($\seq_1$, $\seq_2$, \var{output}):
performs the intersection/union between $\seq_1$ and $\seq_2$,
materializing the result into the \var{output} buffer
of 32-bit integers and returning the size of the result;

\item $\seq$.\textsf{access}($i$):
returns the integer $\seq[i]$;

\item $\seq$.\textsf{nextGEQ}($x$):
returns the integer greater-than or equal-to $x$
(this operations is more classically known as \emph{successor}),
that is the smallest integer $z \geq x$.
If $x$ is larger than the largest element of $\seq$, a default
value is returned, here assumed to be called \var{limit} and
such that $\var{limit} \geq u$.

\end{itemize}

\newpage
Except for the operations \textsf{decode} and \textsf{OR} that
need to sequentially scan the sequence,
an efficient implementation of the aforementioned operations
relies on \emph{partitioning} the sequence
because of the following simple observation:

\return
\emph{When we ask whether the integer $x$ is present or not
in the sequence $\seq$, we can safely skip all partitions of $\seq$ whose maximum
integer is less than $x$ because $\seq$ is sorted, thus none of the integers
less than $x$ should be considered.}

\begin{figure}
\centering
\subfloat[partitioning by cardinality -- PC]{
	\includegraphics[scale=0.75]{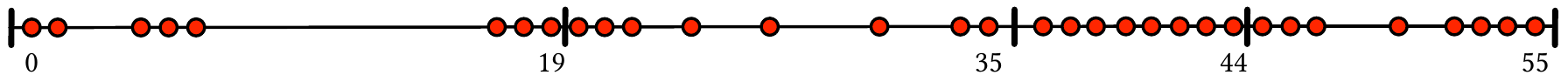}
	\label{fig:c}
}

\subfloat[partitioning by universe -- PU]{
	\includegraphics[scale=0.75]{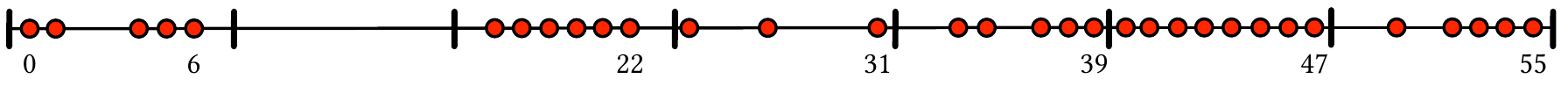}
	\label{fig:u}
}

\caption{Example of a sequence of 32 values drawn from a universe of size 56 as partitioned by cardinality (a) and by universe (b), using partitions of 8 integers. We also mark the maximum integer in each
partition.
\label{fig:partitionings}}
\end{figure}

\return
Classically, integer sequences have been partitioned \emph{by cardinality},
i.e., consecutive elements are grouped together into fixed-size or
variable-size partitions.
However, partitioning a sequence \emph{by universe} is also possible.
In a simple implementation of the approach,
a universe \emph{span} $s$ is chosen and all integers falling
into the $k$-th bucket $[sk, s(k+1))$ are compressed
into the same partition.

Fig.~\ref{fig:partitionings} shows an example of such
paradigms applied to an example sequence
$\seq(32,55) = \langle$0, 1, 4, 5, 6, 17, 18, 19, 20, 21, 22, 24, 27, 31, 34, 35, 37, 38, 39, 40, 41, 42, 43, 44, 45, 46, 47, 50, 52, 53, 54, 55$\rangle$, for partitions of size 8.
In Fig.~\ref{fig:c}, 8 consecutive integers are packed
together, thus the following $32/8 = 4$ partitions are defined
$\langle$0, 1, 4, 5, 6, 17, 18, 19$\rangle$
$\langle$20, 21, 22, 24, 27, 31, 34, 35$\rangle$
$\langle$37, 38, 39, 40, 41, 42, 43, 44$\rangle$
$\langle$45, 46, 47, 50, 52, 53, 54, 55$\rangle$.
In Fig.~\ref{fig:u}, 8 consecutive universe values are packed
together, thus the following $\lceil 55/8 \rceil = 7$ partitions are defined
$\langle$0, 1, 4, 5, 6$\rangle$
$\langle\rangle$
$\langle$17, 18, 19, 20, 21, 22$\rangle$
$\langle$24, 27, 31$\rangle$
$\langle$34, 35, 37, 38, 39$\rangle$
$\langle$40, 41, 42, 43, 44, 45, 46, 47$\rangle$
$\langle$50, 52, 53, 54, 55$\rangle$.
Notice how, in this latter example, partitions may have
different cardinalities and that some of them may be
empty indeed as it happens for the second one
spanning the universe slice $[8,16)$.

\return
The key point is that literature on inverted index compression
has extensively adopted the partitioning-by-cardinality paradigm (PC),
where little attention has been given to the other paradigm (PU).
As a result of this: (1) no experimental comparison between such
paradigms have been assessed in the setting of inverted index
compression (to the best of our knowledge, only one prior
work~\cite{wang2017experimental}
takes a similar issue into account);
(2) few solutions for the PU paradigm have been designed.

\return
Therefore, after a detailed description of the two paradigms
(Section~\ref{sec:paradigms}),
we: first, design a simple PU solution that is tailored for the exploitation of the clustering property
of inverted lists (Section~\ref{sec:slicing}),
namely the fact that inverted lists are notably known
to feature clusters of very close document identifiers
that can be compressed very well;
then, experimentally compare the advantages and disadvantages
of both paradigms in terms of achieved compression effectiveness
and the efficiency of the operations introduced before
(Section~\ref{sec:experiments}).
We finally summarize the experimental findings and sketch some
promising future directions
(Section~\ref{sec:conclusions}).

\section{Paradigms}\label{sec:paradigms}

As already introduced, we individuate two different paradigms that partition a sequence
to achieve efficient query resolution, as exemplified in Fig.~\ref{fig:partitionings} and described in the following:
\emph{partitioning by cardinality} (PC) and \emph{partitioning by universe} (PU).

\begin{figure}[t]

    \subfloat[]{
        \scalebox{0.9}{
            \begin{minipage}{.5\textwidth}
            \input{intersection1.tex}
            \end{minipage}
        }
    }
    \subfloat[]{
        \scalebox{0.9}{
            \begin{minipage}{.5\textwidth}
            \input{intersection2.tex}
            \end{minipage}
         }
    }

  \caption{Two different intersection algorithms, respectively suitable for
  sequences that are partitioned by cardinality (a) and by universe (b).
  In the pseudo code (a), the function $\seq.\func{next}()$
  returns the integer that follows the last returned during a sequential
  scan of $\seq$.
  In pseudo code (b): $\seq.\func{begin}()$ and $\seq.\func{end}()$ returns
  iterators over the partitions of $\seq$, respectively at the beginning
  and at the end;
  $\var{it}.\func{id}()$ returns the identifier of the partition;
  $\var{it}.\func{next}()$ advances iterator \var{it} to the
  next partition; $\var{it}.\func{advance}(\var{id})$ advances iterator
  \var{it} to the first partition whose identifier compares greater-than
  or equal-to \var{id};
  the function $\var{n} = \func{AND}(\var{l}, \var{r}, \var{output})$
  intersects the partitions pointed to by the iterators \var{l} and \var{r},
  writing the result in \var{output} and returning the size \var{n} of the
  intersection.
  \label{alg:intersections}}
\end{figure}

\subsection{Partitioning by cardinality}\label{subsec:card}
Traditionally, an inverted list is \emph{partitioned by cardinality}, i.e.,
consecutive integers of the list are grouped together into a partition
until a given cardinality is not reached.
The cardinality can be \emph{fixed}
for every partition (besides the last one which may contain less integers),
e.g., 128 integers, or can vary according to the actual values of the integers
being compressed, e.g., in order to achieve a more compact representation~\cite{2014:ottaviano.venturini,TKDE19}.

The list also stores the (sorted) sequence formed by the
maximum values of every partition.
The values in such sequence are usually called \emph{skip pointers}.
Skip pointers add a small space overhead to the list representation
itself for reasonably-large cardinality values, but allow skipping over
the inverted list's values.
Such list organization relies on the operation $\seq.\Nextgeq(x)$
to support efficient list intersection.
We can implement $z = \seq.\Nextgeq(x)$ by first searching $x$ within
the skip pointers to individuate the partition where the wanted value $z$ lies in
and, then, conclude the search for it in that partition only.
The operation is efficient because the set of skip pointers is small and
searching a value in a single partition is faster than searching it in the
whole list without any positional restriction.

\return
Let us now consider how list intersection can be achieved through the {\Nextgeq} primitive.
Suppose we have to compute the intersection between the inverted lists associated
to terms $t_1$ and $t_2$, i.e., $\textsf{AND}(t_1,t_2) = \seq_1 \cap \seq_2$,
where $\seq_1$ is shorter than $\seq_2$.
We search for the first value $x$ of $\seq_1$ in $\seq_2$
with $\seq_2.\Nextgeq(x)$: if the value $z$ returned by the operation is equal to $x$
then it is a member of the intersection and we can just repeat this search step
for the \emph{next} value of $\seq_1$;
otherwise $z$ gives us a \emph{candidate} value
to be searched next,
indeed allowing to skip the searches for all values
between $x$ and $z$.
In fact, since we have that $z \geq x$, $\seq_2.\Nextgeq(y)$ will be equal to $z$
also for \emph{all} values $y$ such that $x < y < z$,
thus none of such integers
can be a member of the intersection.
Fig.~\ref{alg:and_a} illustrates such procedure.

\subsection{Partitioning by universe}\label{subsec:univ}
Another strategy is partitioning an inverted list \emph{by universe}, i.e.,
all integers $a \leq x < b$ belong to the same universe-aligned partition $[a,b)$.
For example, given $\seq(n,u)$ we may choose a universe \emph{span} of $s$ integers,
so that the following universe-aligned partitions are defined:
$[0,s)$ $[s, 2s)$ $[2s, 3s)$ $\cdots$ $[\lfloor u/s \rfloor, u)$.
Note that such partitions do \emph{not} depend on the actual size $n$ of the list,
but only on its universe of representation.
All integers less than $s$ are grouped into the first partition; all integers
less than $2s$ and larger than or equal to $s$ are grouped into the second partition; and so on.
In general, if the $k$-th partition contains $m$ integers, i.e.,
there are $m$ integers $x$ such that $sk \leq x < s(k+1)$, we say that
the partition has \emph{cardinality} equal to $m \leq s$.

Also this strategy permits to skip over the list values because only
partitions relative to the \emph{same} universe have to be intersected.
Now the skip pointers are represented by enumerating
the non-empty partitions, rather than being actual list values.
Therefore, list intersection proceeds by identifying all common partitions,
and, for each of them, by resolving a smaller intersection of at most $s$ integers.
Depending on the actual cardinality of a partition, different compression
strategies AND/OR intersection algorithms can be adopted, thus
\emph{not necessarily}
relying on the {\Nextgeq} primitive.
Fig.~\ref{alg:and_b} illustrates this other approach.

On the other hand, the overhead represented by the skip pointers may be excessive
for very sparse inverted lists, because the partitions do not depend on the value of $n$.
In fact, in the worst case, we could be maintaining
a pointer for each integer in the list (each partition
is a singleton).

\return
As already claimed, this paradigm has been used in the coding
and (theoretical) data structure design areas.
We briefly discuss some old, but yet very meaningful, examples:
the Elias-Fano encoding algorithm~\cite{Fano71,Elias74}
and the van Emde Boas data structure~\cite{Boas75,Boas77}.

Elias-Fano represents a sequence $\seq(n,u)$ in at most
$n\phi + n + \lceil u/2^{\phi} \rceil$ bits.
It can be shown~\cite{Elias74} that choosing
$\phi = \lfloor \log_2\frac{u}{n} \rfloor$ minimizes the
number of bits. In other words, the values of $\seq$
are partitioned by universe into chunks containing at most $2^{\phi}$
integers each.
We refer to this split as \emph{parametric}, because it
is dependent on the value of $u$ \emph{and} $n$, thus making
the intersection algorithm shown in Fig.~\ref{alg:and_b}
not directly applicable because sequences having different
sizes partition the integers differently.

But a non-parametric split is possible as well.
For example, assuming a universe of size
$u = 2^{32}$, we could
partition $u$ into $\sqrt{u} = 2^{16}$ chunks, i.e.,
$\phi = 16$.
In this way, each chunk contains all integers
sharing the same 16 most significant bits.
This is reminiscent of the van Emde Boas data structure:
a recursive tree layout
solving the well-known \emph{dictionary problem}
(see the introduction to parts III and V of the
book by~\citet{CLRS}) in $O(\log\log u)$ time per operation
and $O(u)$ words of space\footnote{Actually, the space can be improved to
$O(n)$ words using \emph{bucketing}. See this blog post by
Mihai P\v{a}tra\c{s}cu: \url{http://infoweekly.blogspot.com/2007/09/love-thy-predecessor-iii-van-emde-boas.html}.}.
In such data structure, the universe $u$ is recursively
partitioned into $\sqrt{u}$ chunks.

A similar fixed-universe partitioning approach as been recently adopted by
\emph{Roaring}~\cite{chambi2016better,lemire2016consistently,lemire2018roaring},
a practical data structure
that has been shown to outperform all previously proposed
bitmap indexes~\cite{chambi2016better,wang2017experimental}
and it is widely used in commercial applications.
Specifically, Roaring partitions $u$ into chunks of $2^{16}$ integers and
represents all the integers falling into a chunk
in two different ways according to the cardinality of the chunk:
if a chunk contains less than 4096 elements,
then it is represented as a sorted array
of 16-bit integers; otherwise it is represented as a bitmap of $2^{16}$ bits.
Finally, extremely dense chunks can also be represented with runs.
For example, the two runs $(13,42)(60,115)$ mean that all
the integers
$13 \leq x \leq 13+42$ and
$60 \leq x \leq 60+115$ belong to the chunk.

\section{The Slicing approach}\label{sec:slicing}

In this section we design a solution that applies a recursive
universe slicing approach to achieve compact storage
and good practical
performance for the operations introduced
in Section~\ref{sec:introduction}.

Let us consider a strictly increasing sequence $\seq(n,u)$
whose elements are drawn from a universe of size $u \leq 2^{32}$.
At a high-level point of view, we represent $\seq$
using a tree of height 3 (at most), where the root
has fanout $s_1$ and its children have fanout $s_2$.
Refer to Fig.~\ref{fig:tree}.
We now detail how the data structure is concretely
implemented and operations supported.

\subsection{Data structure}
The root of the tree logically corresponds
to the interval $[0,u)$ that is partitioned into
slices spanning $s_1$ integers each
(except, possibly, the last slice which may contain less integers).
In what follows, we refer to such $s_1$-long
slices as \emph{chunks}.
A header array $H_1$ is used to classify chunks into
4 different types according to their cardinality:
full, dense, sparse and empty.
Full chunks, i.e., containing exactly $s_1$ integers,
and empty chunks (containing no integers at all)
are represented implicitly by their types.
A dense chunk spanning the $k$-th universe slice
$[s_1 k, s_1 (k+1))$ is represented with a bitmap of $s_1$ bits
by setting the $i$-th bit if the integer $i - s_1 k$ belongs
to the slice.
In particular, we regard a chunk to be dense if its cardinality
is at least $s_1/2$.
Doing so guarantees that the average number of bits
spent for each integer belonging to a dense chunk is at most 2.
Therefore, full, empty and dense chunks have no children.
Instead, sparse chunks are encoded by re-applying the same strategy:
since every sparse chunk now logically corresponds to
a (smaller) universe slice of size $s_1$,
the interval $[0,s_1)$ is partitioned into slices
spanning
$s_2$ integers each
(again, except possibly the last one).
We refer to such $s_2$-long
slices as \emph{blocks}.
As before, a header array $H_2$ is used to distinguish
between different block types.
However, given the smaller universe slice,
we only distinguish between two block types,
namely dense and sparse,
in order to better amortize the cost of $H_2$.
In particular, a dense block is encoded with a bitmap of $s_2$ bits;
a sparse block is represented with a sorted array of
$\lceil \log_2 s_2 \rceil$-bit integers.

\begin{figure}
\centering
\subfloat[tree-like view]{
	\includegraphics[scale=0.75]{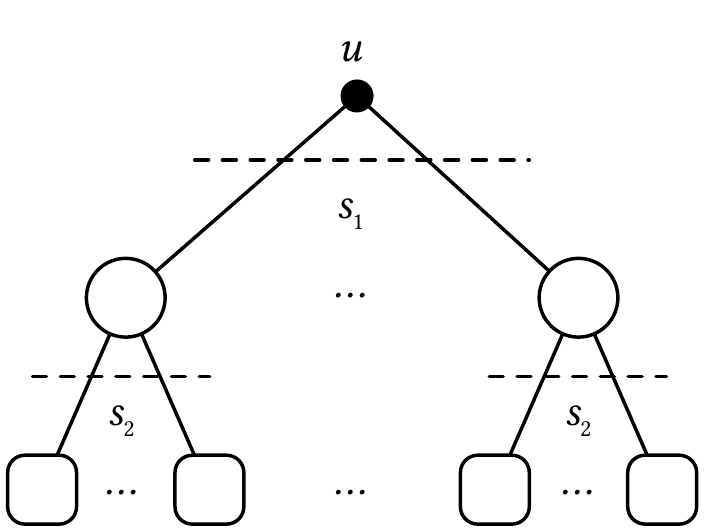}
	\label{fig:tree}
}

\subfloat[array-like view]{
	\includegraphics[scale=0.62]{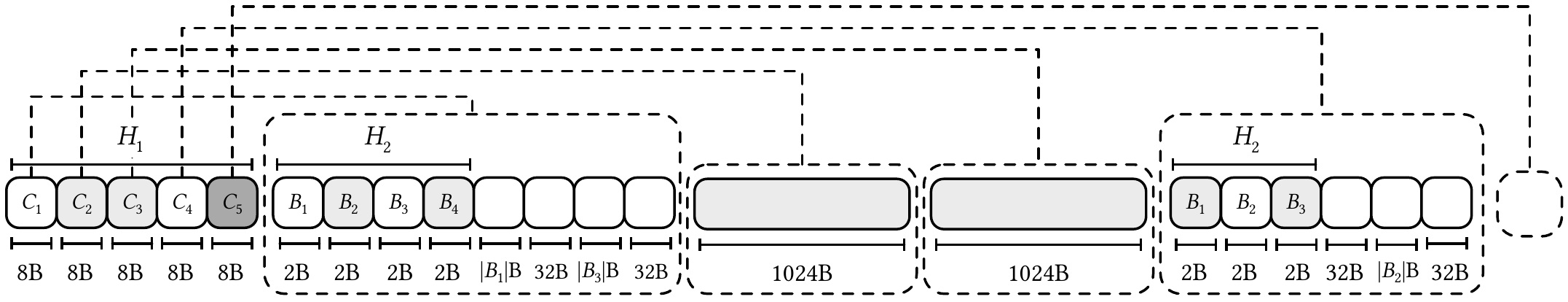}
	\label{fig:example}
}

\caption{The upper part (a) of the picture shows the recursive
slicing of the universe $u$ made by the data structure, here viewed
as a tree.
In the lower part (b), we show
how the data structure is concretely laid out
in memory, for an example sequence made up of chunks $C_1..C_5$.
In this example, chunks $C_2$ and $C_3$ are dense, hence represented
with bitmaps of 1024 bytes (B); $C_1$ and $C_4$ are sparse, thus
recursively decomposed; lastly, $C_5$ is full, hence implicitly
represented. Below each box -- corresponding to a universe slice --
we indicate the number of bytes taken
by its encoding. Dashed lines help to graphically translate the
array-like view into the tree-like view.}
\end{figure}

\return
We choose $s_1 = 2^{16}$ and $s_2 = 2^8$.
Since we consider the case when $u \leq 2^{32}$ and for
our choice of $s_1 = 2^{16}$, the number of chunks is always
at most $2^{16}$, thus we encode this quantity into 16 bits.
Similarly, it follows that each sparse chunk
is sliced into at most $2^8$ blocks.
With this choice of $s_1$ and $s_2$, a dense chunk is
a bitmap of 1024 bytes; a dense block is a bitmap of 32 bytes;
a sparse block whose cardinality is $c$ is a sorted array
of 8-bit integers, hence taking $c$ bytes overall.

For each \emph{non-empty} block $[k 2^8, (k+1) 2^8)$,
the array $H_2$ stores its identifier $k$ and its cardinality.
Both quantities are always at most $2^8$, thus they take one byte
each. Knowing the cardinality $0 < c \leq 2^8$
of a block we derive the number
of bytes needed by its representation.
If $c < 2^8/8 - 1 = 31$, the block is considered to be sparse,
thus it takes $c$ bytes; otherwise it is dense and takes 32 bytes.
Therefore, a sparse block consumes at most $(2^8/8 - 2)\times 8 + 8 = 248$ bits.
Note that we do not set the sparseness threshold to $2^8/8 = 32$ because
otherwise a sparse block would consume at most $256$ bits
that is equal to the cost of a dense block and a bitmap would suffice.

For each \emph{non-empty} chunk $[k 2^{16}, (k+1) 2^{16})$,
the array $H_1$ stores, instead, the following quantities:
its identifier $k$, its cardinality, and
the number of bytes needed by its encoding.
Similarly to the case of a sparse block,
we require the encoding of a sparse chunk to take
\emph{less than} $2^{16}$ bits, otherwise a bitmap of $2^{16}$
bits would suffice.
Therefore, each of these 3 quantities easily fits into a
16-bit integer.

Although we could derive the type of a chunk from its
cardinality as done for a block,
we also store the type explicitly using 16 bits.
When a chunk is sparse, we need to know the number
of its blocks, i.e., the number of non-empty $2^8$-size slices.
This number is the size of the corresponding header $H_2$.
Therefore, we write this quantity in 8 bits and interleaved with
the 16 bits dedicated to the type information.
In conclusion, we spend a 64-bit overhead per chunk.

\return
Fig.~\ref{fig:example} shows an example of such organization and
how all the different data quantities (headers, bitmaps and arrays)
are laid out in memory.
In practice, the logical tree shown in Fig.~\ref{fig:tree} is
``linearized'' into an array of bytes.

\return
The data structure described here
has similarities with some previous approaches.
As already discussed, the universe is exponentially reduced
like in a van Emde Boas tree,
i.e., $2^{32} \rightarrow 2^{16} \rightarrow 2^8$.
Partitioning the universe recursively has the potential of
adapting to the distribution of the integers being encoded,
a crucial design choice for clustered integer sequences
such as inverted lists.
The choice of bitmaps to represents dense sets is a widely
adopted technique, employed by, for example,
partitioned Elias-Fano~\cite{2014:ottaviano.venturini},
hybrid Variable-Byte schemes~\cite{TKDE19}
and Roaring~\cite{lemire2018roaring}.
However, as we are going to show, the use of bitmaps joint
with universe-aligned partitions is particularly effective
for fast query execution because operations
can be implemented via
inexpensive bitwise instructions, hence exploiting
word-level parallelism, and are suitable
for even more advanced instructions, such as SIMD AVX.

The description above also opens the possibility
for better compression.
For example, we could use a different representation for
sparse blocks, e.g., bit-aligned universal codes.
Whatever representation
we use, that will give birth to
interesting time/space trade-offs.
The choice adopted here of $s_1 = 2^{16}$,
$s_2 = 2^8$ and the use of
8-bit integer arrays clearly favours time efficiency given
that both bitmaps and packed arrays are aligned to
byte boundaries.

\subsection{Operations}\label{subsec:operations}
We now describe how the operations are supported by the
data structure.

\parag{Decoding}
The $\seq$.\textsf{decode}(\var{output}) operation
decodes $\seq$ sequentially to the \var{output} buffer
of 32-bit integers. We loop through each chunk and, depending
on its type, we decode it accordingly appending the
result into the \var{output} buffer.
This permits to write different specialized functions
to handle a slice differently based on its type.

Bitmaps can be efficiently decoded using the built-in function
\texttt{ctzll} which counts the number of
trailing zero into a 64-bit word~\cite{lemire2018roaring}
(we also tested a SIMD algorithm to decode larger bitmaps but
got almost no speed improvement).

The sorted arrays encoding sparse blocks
contain at most $2^8/8-2$ integers, each value
taking 8 bits.
Therefore, we can use the
SIMD instruction \texttt{\_mm256\_cvtepu8\_epi32},
that zero extend packed
(unsigned) 8-bit integers to 32-bit integers.
Doing so, we can efficiently decode 8 values at a time.
We also observed that this approach is even more efficient
when paired with loop unrolling, thus we apply the instruction either
2 or 4 times after a single test on the block cardinality.

\parag{Intersection}
The operation \textsf{AND}($\seq_1$, $\seq_2$, \var{output})
performs the intersection between $\seq_1$ and $\seq_2$,
materializing the result into the \var{output} buffer
of 32-bit integers and returning the size of the result.
We use the algorithm illustrated in Fig.~\ref{alg:and_b},
thus we loop through the header arrays of the sequences,
intersecting only chunks/blocks that are shared by the two
(line 8).
Therefore, we reduce the problem of list intersection to the
smaller instance of performing intersections between
(1) two bitmaps, or (2) two arrays,
or a (3) bitmap and an array
(line 9).

Case (1) -- the intersection between two bitmaps -- translates into a sequence
of inexpensive bitwise
\textsf{AND} instructions between 64-bit words with
(usually) automatic compiler vectorization.

\begin{figure}
\begin{lstlisting}
#define INIT                                                \
    __m256i base_v = _mm256_set1_epi32(base);               \
    __m128i v_l = _mm_lddqu_si128((__m128i const*)l); \
    __m128i v_r = _mm_lddqu_si128((__m128i const*)r); \
    __m256i converted_v;                                    \
    __m128i shuf, p, res;                                   \
    int mask, matched;

#define INTERSECT                                                             \
    res =                                                                     \
        _mm_cmpestrm(v_l, c_l, v_r, c_r,                                      \
                     _SIDD_UBYTE_OPS |                                        \
                     _SIDD_CMP_EQUAL_ANY |                                    \
                     _SIDD_BIT_MASK);                                         \
    mask = _mm_extract_epi32(res, 0);                                         \
    matched = _mm_popcnt_u32(mask);                                           \
    size += matched;                                                          \
    shuf = _mm_load_si128((__m128i const*)T + mask);                          \
    p = _mm_shuffle_epi8(v_r, shuf);                                          \
    converted_v = _mm256_cvtepu8_epi32(p);                                    \
    converted_v = _mm256_add_epi32(base_v, converted_v);                      \
    _mm256_storeu_si256((__m256i*)out, converted_v);                          \
    if (matched > 8) {                                                        \
        p = _mm_bsrli_si128(p, 8);                                            \
        converted_v = _mm256_cvtepu8_epi32(p);                                \
        converted_v = _mm256_add_epi32(base_v, converted_v);                  \
        _mm256_storeu_si256((__m256i*)(out + 8), converted_v);                \
    }

#define ADVANCE(ptr)                                     \
    out += size;                                         \
    ptr += 16;                                           \
    v_##ptr = _mm_lddqu_si128((__m128i const*)ptr); \
    c_##ptr -= 16;
\end{lstlisting}
\caption{C++ macros to support the vectorized implementation
of the intersection between small sorted arrays.
These macros are used by the code shown in Fig.~\ref{alg:and_SIMD_complete}.
\label{alg:and_SIMD}}
\end{figure}

Case (2) has to intersect tiny 8-bit sorted arrays.
While a scalar textbook intersection algorithm between
uncompressed arrays would suffice, we can accelerate the
process using a variation of
the vectorized approach by~\citet*{schlegel2011fast}.
In short, the algorithm uses the
SIMD instruction \texttt{\_mm\_cmpestrm}
to compare strings of bytes.
In our case we can, therefore,
execute an all-versus-all comparison in parallel between
sets of 16 $\times$ 8-bit integers.
Matching integers, i.e., integers in common between the two
sets, are marked with a 32-bit bitmap returned as the result
of the comparison.
We can use this 32-bit value as an index in a pre-computed
universal table $T$ of 1024 $\times$ 1024 bytes
to obtain a permutation of bytes indexes,
indicating how the matching integers should be permuted
to collate them to the beginning of a
128-bit register.
Such permutation is applied with the dedicated 
\texttt{\_mm\_shuffle\_epi8} SIMD instruction.
The C++ macro \texttt{INTERSECT} shown in Fig.~\ref{alg:and_SIMD}
illustrates this approach.
Let $c_l$ and $c_r$ be the cardinality of the two sets respectively.
Since in our case we have that both $c_l$ and $c_r$
are less than 32,
we can directly enumerate the following 3 different cases:
(1) $c_l \leq 16$ and $c_r \leq 16$, then we need only 1
string comparison;
(2) $c_l \leq 16$ and $c_r > 16$ (or $c_r \leq 16$ and $c_l > 16$),
then we need 2 string comparisons;
(3) $c_l > 16$ and $c_r > 16$, then we would need 4 string
comparisons but we determined that the simple scalar version
is more efficient.
The C++ function \texttt{sparse\_blocks\_and} coded in
Fig.~\ref{alg:and_SIMD_complete}
shows these cases and, along with the code in
Fig.~\ref{alg:and_SIMD}, completes our intersection algorithm for
small sets.

For detailed descriptions of SIMD instructions, refer to the
excellent Intel guide at \url{https://software.intel.com/sites/landingpage/IntrinsicsGuide}.

\begin{figure}
\begin{lstlisting}
size_t sparse_blocks_and(uint8_t const* l, uint8_t const* r,
                           int c_l, int c_r,
                           uint32_t base, uint32_t* out) {
    size_t size = 0;
    if (c_l <= 16 and c_r <= 16) {
        INIT INTERSECT return size;  // 1 cmpestr
    }
    if (c_l <= 16 and c_r > 16) {
        INIT INTERSECT ADVANCE(r) INTERSECT return size;  // 2 cmpestr
    }
    if (c_r <= 16 and c_l > 16) {
        INIT INTERSECT ADVANCE(l) INTERSECT return size;  // 2 cmpestr
    }
    /* scalar code goes here */
}
\end{lstlisting}
\caption{The C++ skeleton of the vectorized implementation
of the intersection between small sorted arrays.
The macros \texttt{INIT}, \texttt{INTERSECT} and \texttt{ADVANCE}
are shown in Fig.~\ref{alg:and_SIMD}.
In the code \texttt{c\_l} and \texttt{c\_r} indicate
the cardinalities of the two arrays respectively;
the value \texttt{base} is equal to
$k_1 2^{16} + k_2 2^{8}$ if we are intersecting
the $k_2$-th $2^{8}$-long slices of the
$k_1$-th $2^{16}$-long slice.
\label{alg:and_SIMD_complete}}
\end{figure}

Case (3) -- the intersection between a bitmap and an array --
is implemented by checking if the values of the
array correspond to bits set in the bitmap,
using the bit-test assembler instruction.

\parag{Union}
The operation \textsf{OR}($\seq_1$, $\seq_2$, \var{output})
performs the union between $\seq_1$ and $\seq_2$,
materializing the result into the \var{output} buffer
of 32-bit integers and returning the size of the result.
The algorithm follows the same skeleton described for
the intersection, albeit we do not rely on specific
SIMD optimizations: bitmaps are merged using bitwise \textsf{OR}
within 64-bit words;
sorted arrays using scalar code;
the case with a bitmap and an array is handled by first
converting the sorted array into a bitmap, then using
the parallelism of bitwise \textsf{OR}.

\parag{Random access}
The operation $\seq$.\textsf{access}($i$)
returns the integer $\seq[i]$.
We scan the header array of the data structure to
take into account for the cardinality of each chunk
covering a universe of size $2^{16}$ in order to
locate the chunk containing the $i$-th integer.
To make this search faster,
we build cumulative cardinality counts for groups of $a$ non-empty
universe chunks, thus skipping $a$ chunks if the sum of their cardinalities is less than $i$.
The parameter $a$ is an \emph{associativity} value that in our implementation we set to 32.
Then we proceed recursively at the block-level
if a chunk is sparse
(but we do not build cumulative counts at the block level).

In particular, whenever we encounter a bitmap, we rely on
efficiency of the built-in instruction
\texttt{popcountll}
to locate the 64-bit word where the wanted integer lies in.
This instruction returns the number of bits set
in a 64-bit word.
Now that we have reduced the problem to a word of 64 bits,
we can use the parallel-bit deposit assembler
instruction \texttt{pdep} to perform a fast select-in-word
operation~\cite{pandey2017fast}.

\parag{nextGEQ}
The operation $\seq$.\textsf{nextGEQ}($x$)
returns the integer greater-than or equal-to $x$,
that is the smallest integer $z \geq x$.
Since our data structure is partitioned by universe,
we can directly identify the chunk comprising $z$
because this is the one having identifier $x / 2^{16}$,
i.e., we consider the 16 most significant bits of the
key $x$. The wanted value $z$ lies in such
partition or, if $x$ is larger than the maximum
value in the partition, it is the minimum (first)
value in the partition that follows.
Observe that this operation is actually faster than {\access}
for universe-aligned methods, because it does not
need to \emph{search} for the wanted partition.

\section{Experiments}\label{sec:experiments}

The aim of this section is twofold: establishing
a solid experimental comparison between the two
different paradigms described in Section~\ref{sec:paradigms}
in order to assess the achievable space/time trade-offs
and reporting on the effectiveness/efficiency of the
Slicing approach introduced in Section~\ref{sec:slicing}.

\parag{Tested configurations}
We compare the configurations summarized in
Table~\ref{tab:strategies}
for the following reasons.
For the paradigm \emph{partitioning by cardinality} with
\emph{fixed}-sized partitions of 128 integers, we test:
Variable-Byte~\cite{thiel1972program}
with the SIMD-ized decoding algorithm devised by~\citet{2015:plaisance.kurz.ea};
Interpolative~\cite{2000:moffat.stuiver}
and Elias-Fano~\cite{2014:ottaviano.venturini} as representative of,
respectively, highest speed, best
compression effectiveness and best space/time trade-off in the literature.
As representative of the paradigm \emph{partitioning by cardinality} with
\emph{variable}-sized partitions, we test the $\epsilon$-optimal Elias-Fano
mechanism~\cite{2014:ottaviano.venturini}.
For all such representation, we use the C++ implementation provided in the
\texttt{ds2i} library, available at \url{https://github.com/ot/ds2i}.

Concerning the paradigm \emph{partitioning by universe}, we test three solutions.
The first two solutions are represented by {Roaring}~\cite{lemire2018roaring} (see Section~\ref{subsec:univ}).
We test the solution without the run-container optimization, thus
using two container types (bitmap and sorted array), and with the
optimization, thus using three container types
(bitmap, sorted array and
run).
We use the dedicated library written in C and available at \url{https://github.com/RoaringBitmap/CRoaring}.

The third solution is the {Slicing} approach described
in Section~\ref{sec:slicing}.
Our C++ implementation of the mechanism is freely available at
\url{https://github.com/jermp/s_indexes}.

\begin{table}
\centering
\scalebox{\mytablescale}{\begin{tabular}{l c l }
\toprule

Method & Shorthand & Strategy
\\

\cmidrule(lr){1-1}
\cmidrule(lr){2-2}
\cmidrule(lr){3-3}

Variable-Byte & {\vb}
& {PC; fixed-sized partitions of 128 integers; byte-aligned}
\\

Elias-Fano & {\uniform}
& {PC; fixed-sized partitions of 128 integers; bit-aligned}
\\

Interpolative & {\interp}
& {PC; fixed-sized partitions of 128 integers; bit-aligned}
\\


Elias-Fano $\epsilon$-opt. & {\opt}
& {PC; variable-sized partitions; bit-aligned}
\\


Roaring without run opt. & {\roar}
& {PU; single-span; 2 container types; byte-aligned}
\\

Roaring with run opt. & {\roaropt}
& {PU; single-span; 3 container types; byte-aligned}
\\

Slicing & {\slice}
& {PU; multi-span; byte-aligned}
\\

\bottomrule
\end{tabular}}
{\aftertabspace}
\caption{The different tested configurations.}
\label{tab:strategies}
\end{table}

\parag{Datasets} We perform the experiments on the following standard test collections.

\begin{itemize}
\item {\gov} is the TREC 2004 Terabyte Track test collection,
consisting in roughly 25 million \textsf{.gov} sites crawled in early 2004.
The documents are truncated to 256 KB.
\item {\clue} is the ClueWeb 2009 TREC Category B test collection,
consisting in roughly 50 million English web pages
crawled between January and February 2009.
\item {\cc} is a dataset of news freely available from
CommonCrawl: \url{http://commoncrawl.org/2016/10/news-dataset-available}.
Precisely, the datasets consists of the news appeared from 09/01/16 to 30/03/18.
\end{itemize}

\noindent
Identifiers were assigned to documents according to the lexicographic order
of their URLs~\cite{2007:silvestri}.
Table~\ref{tab:datasets} reports the basic statistics for the collections.
We choose three different
levels of list density $d$, i.e., the ratio between the size of a list and its maximum integer, and compress all lists whose density exceeds $d$.
By varying the density, we highlight
how compression effectiveness changes for the two different
partitioning paradigms used,
still focusing on \emph{most} of the integers in the collections.
Refer to Table~\ref{tab:densities}.

\begin{table}
\centering
\scalebox{\mytablescale}{\begin{tabular}{l r r r}
\toprule

Statistic & {\gov} & {\clue}  & {\cc} \\

\cmidrule(lr){1-1}
\cmidrule(lr){2-2}
\cmidrule(lr){3-3}
\cmidrule(lr){4-4}

Sequences
  & \num{35636425}
  & \num{92094694}
  & \num{43844574}
  \\
  
Universe
  & \num{24622347}
  & \num{50131015}
  & \num{43530315}
  \\

Integers
  & \num{5742630292}
  & \num{15857983641}
  & \num{20150335440}  
  \\
  
\bottomrule
\end{tabular}}
{\aftertabspace}
\caption{Basic statistics for the test collections.}
\label{tab:datasets}
\end{table}

\begin{table}
\centering
\scalebox{\mytablescale}{\begin{tabular}{c l r r r}
\toprule

Density & Statistic & {\gov} & {\clue} & {\cc}
\\

\cmidrule(lr){1-1}
\cmidrule(lr){2-2}
\cmidrule(lr){3-3}
\cmidrule(lr){4-4}
\cmidrule(lr){5-5}

\multirow{3}{*}{$10^{-2}$}
& Sequences
  & \num{3513}
  & \num{5802}
  & \num{5930}
  \\

& Integers
  & \num{4347653438}
  & \num{11676154022}
  & \num{16677342102}   
  \\

& \%
  & 76
  & 74
  & 83
  \\

\cmidrule(lr){1-1}
\cmidrule(lr){2-2}
\cmidrule(lr){3-3}
\cmidrule(lr){4-4}
\cmidrule(lr){5-5}

\multirow{3}{*}{$10^{-3}$}
& Sequences
  & \num{13276}
  & \num{21924}
  & \num{23085}
  \\

& Integers
  & \num{5066748826}
  & \num{13864451283}
  & \num{18969946075}   
  \\

& \%
  & 88
  & 87
  & 94
  \\

\cmidrule(lr){1-1}
\cmidrule(lr){2-2}
\cmidrule(lr){3-3}
\cmidrule(lr){4-4}
\cmidrule(lr){5-5}

\multirow{3}{*}{$10^{-4}$}
& Sequences
  & \num{85893}
  & \num{99227}
  & \num{79954}
  \\

& Integers
  & \num{5390038277}
  & \num{14805194135}
  & \num{19681352639}  
  \\

& \%
  & 94
  & 93
  & 98
  \\
  
\bottomrule
\end{tabular}}
{\aftertabspace}
\caption{Dataset statistics for three levels of density. We also indicate the percentage
of integers retained from the original collections shown in Table~\ref{tab:datasets}.}
\label{tab:densities}
\end{table}

\parag{Experimental setting, methodology and testing details}
The experiments are performed on a machine with 4 Intel
i7-4790K CPUs clocked at 4.00 GHz, with 32 GB of RAM DDR3 and 
running Linux 4.13.0.
All the code is compiled with \textsf{gcc} 7.2.0 using the highest optimization setting
(compilation flags \texttt{-march=native} and \texttt{-O3}).

For the \texttt{CRoaring} library, we compile the code as recommended in the
documentation for best performance,
i.e., with full support for vectorization.
The run-container optimization is enabled
by calling the \texttt{run\_optimize} function.
The implementations of Elias-Fano (both with fixed and variable partitions)
and Interpolative do not use explicit vectorization; the implementation
of Variable-Byte makes use of the vectorized algorithm
devised by~\citet{2015:plaisance.kurz.ea}, called Masked-VByte.

We build the indexes in internal memory and write the corresponding
data structure to a file on disk.
To perform the queries, the data structure is memory mapped from the file
(for \texttt{CRoaring}, by using the \texttt{frozen\_view} function)
and a warming-up run is executed to fetch the necessary pages from disk.

To sequentially decode the indexes, the kernel is also instructed
to access the memory mapped area sequentially using \texttt{posix\_madvice}
with flag \texttt{POSIX\_MADV\_SEQUENTIAL}.
To test the speed of list queries, namely AND/OR, we generated
1000 random pairs of integers and execute the queries with the
corresponding lists.
For point queries, namely {\access} and {\Nextgeq},
we similarly execute 1000 random queries
for each list of the index.
In particular, the 1000 random positions for the {\access} query
are \emph{not} sorted.
The input integers for
{\Nextgeq} are not sorted either and less then the
maximum integer in the sequence (thus, the result is always well determined).

Each run of queries is repeated 10 times
to smooth fluctuations during measurements.
The time reported is the average among these runs.

\begin{table}
\centering
\scalebox{\mytablescale}{\begin{tabular}{c SSS l SSS l SSS}
\toprule

Method
& \multicolumn{3}{c}{$d = 10^{-2}$}
&
& \multicolumn{3}{c}{$d = 10^{-3}$}
&
& \multicolumn{3}{c}{$d = 10^{-4}$}
\\

\cmidrule(lr){1-1}
\cmidrule(lr){2-4}
\cmidrule(lr){6-8}
\cmidrule(lr){10-12}

& {\gov} & {\clue} & {\cc}
&
& {\gov} & {\clue} & {\cc}
&
& {\gov} & {\clue} & {\cc}
\\

\cmidrule(lr){2-2}
\cmidrule(lr){3-3}
\cmidrule(lr){4-4}
\cmidrule(lr){6-6}
\cmidrule(lr){7-7}
\cmidrule(lr){8-8}
\cmidrule(lr){10-10}
\cmidrule(lr){11-11}
\cmidrule(lr){12-12}

{\vb}
& 8.59874 & 8.72134 & 8.66218
&
& 8.71601 & 9.00413 & 9.07528
&
& 8.84851 & 9.18816 & 9.2832
\\

{\uniform}
& 2.72358 & 4.4415 & 4.71734
&
& 3.24897 & 5.14416 & 5.365
&
& 3.65433 & 5.56265 & 5.65584
\\

{\interp}
& 2.33174 & 3.58972 & 4.37232
&
& 2.72138 & 4.11178 & 4.96699
&
& 3.01666 & 4.40728 & 5.23847
\\

{\opt}
& 2.36912 & 4.00844 & 4.5165
&
& 2.84603 & 4.61834 & 5.15798
&
& 3.20086 & 4.96094 & 5.44576
\\

{\roar}
& 6.003608 & 8.881072 & 8.252430
&
& 7.02780 & 9.98834 & 9.21168
&
& 7.601203 & 10.474733 & 9.525385
\\

{\roaropt}
& 5.331765 & 8.494709 & 8.216347
&
& 6.245421 & 9.402477 & 9.169742
&
& 6.745314 & 9.751545 & 9.480298
\\

{\slice}
& 3.23281 & 5.4384 & 5.97909
&
& 3.91038 & 6.39024 & 7.1822
&
& 4.45539 & 7.00042 & 7.76609
\\

\bottomrule
\end{tabular}}
{\aftertabspace}
\caption{Space in bits per integer by varying density $d$.}
\label{tab:space}
\end{table}

\begin{table}
\centering
\scalebox{\mytablescale}{
\begin{tabular}{c SSS l SSS l SSS}
\toprule

Method
& \multicolumn{3}{c}{$d = 10^{-2}$}
&
& \multicolumn{3}{c}{$d = 10^{-3}$}
&
& \multicolumn{3}{c}{$d = 10^{-4}$}
\\

\cmidrule(lr){1-1}
\cmidrule(lr){2-4}
\cmidrule(lr){6-8}
\cmidrule(lr){10-12}

& {\gov} & {\clue} & {\cc}
&
& {\gov} & {\clue} & {\cc}
&
& {\gov} & {\clue} & {\cc}
\\

\cmidrule(lr){2-2}
\cmidrule(lr){3-3}
\cmidrule(lr){4-4}
\cmidrule(lr){6-6}
\cmidrule(lr){7-7}
\cmidrule(lr){8-8}
\cmidrule(lr){10-10}
\cmidrule(lr){11-11}
\cmidrule(lr){12-12}

{\vb}
& 0.513518 & 0.606247 & 0.528292
&
& 0.550443 & 0.662528 & 0.58574
&
& 0.582142 & 0.706762 & 0.618797
\\

{\uniform}
& 0.874415 & 1.28632 & 1.36301
&
& 0.941314 & 1.34128 & 1.41179
&
& 0.980144 & 1.35715 & 1.42167
\\

{\interp}
& 5.26176 & 6.7332 & 7.70933
&
& 5.54049 & 6.94862 & 7.8628
&
& 5.7008 & 7.00913 & 7.90128
\\

{\opt}
& 0.784609 & 1.14684 & 1.34422
&
& 0.857179 & 1.2216 & 1.48046
&
& 0.912888 & 1.25357 & 1.5303
\\

{\roar}
& 0.532040 & 0.716252 & 0.682651
&
& 0.525104 & 0.697586 & 0.685258
&
& 0.537158 & 0.706256 & 0.687055
\\

{\roaropt}
& 0.552781 & 0.755715 & 0.702587
&
& 0.548563 & 0.760792 & 0.690020
&
& 0.573709 & 0.776009 & 0.699855
\\

{\slice}
& 0.563091 & 0.670121 & 0.652603
&
& 0.571514 & 0.68987 & 0.673231
&
& 0.598734 & 0.725803 & 0.706615
\\

\bottomrule
\end{tabular}}
{\aftertabspace}
\caption{Average nanoseconds per decoded integer by varying density $d$.}
\label{tab:decoding}
\end{table}

\parag{Organization}
We organize the experiments in three subsections.
At the whole index level (Section~\ref{sec:space_decoding}), we are interested in the number of bits spent
per represented integer and the time spent per decoded integer when decoding sequentially
every list in the index.
At the list level (Section~\ref{sec:list_queries}), we report the time needed to compute pair-wise
conjunctions (i.e., intersections or boolean AND queries) and pair-wise disjunctions (i.e., unions or boolean OR queries).
Finally, at the single integer level (Section~\ref{sec:point_queries}), we evaluate the time needed to decode an integer at a random position and resolve a {\Nextgeq} query.

\subsection{Index space and decoding time}\label{sec:space_decoding}
Table~\ref{tab:space} reports the average number of bits per integer spent by the different methods.
Clearly, the bit rate is increasing for decreasing
values of density: the sparser a list is, the less
clustered it is, thus more bits are needed to represent 
the values.
In general, across all density levels, the bit-aligned methods {\uniform}, {\opt}
and {\interp} offer the best compression effectiveness,
with the latter being the most space-efficient of all.
Adapting the sizes of the partitions to the distribution
of the integers being compressed pays off: {\opt}
is always more effective than {\uniform}.
The byte-aligned methods {\vb}, {\roar} and {\roaropt} are always
the largest, with {\roar} and {\roaropt} being always more effective than {\vb} on {\gov} but less effective on the other datasets {\clue} and {\cc}.
The use of run containers for the {\roaropt} mechanism pays
off on the more clustered {\gov}, but has a smaller
impact on {\clue} and {\cc}.
In general, between the most effective
methods and the least effective ones there is a factor
of $\approx$2 in space consumption.

\begin{figure}
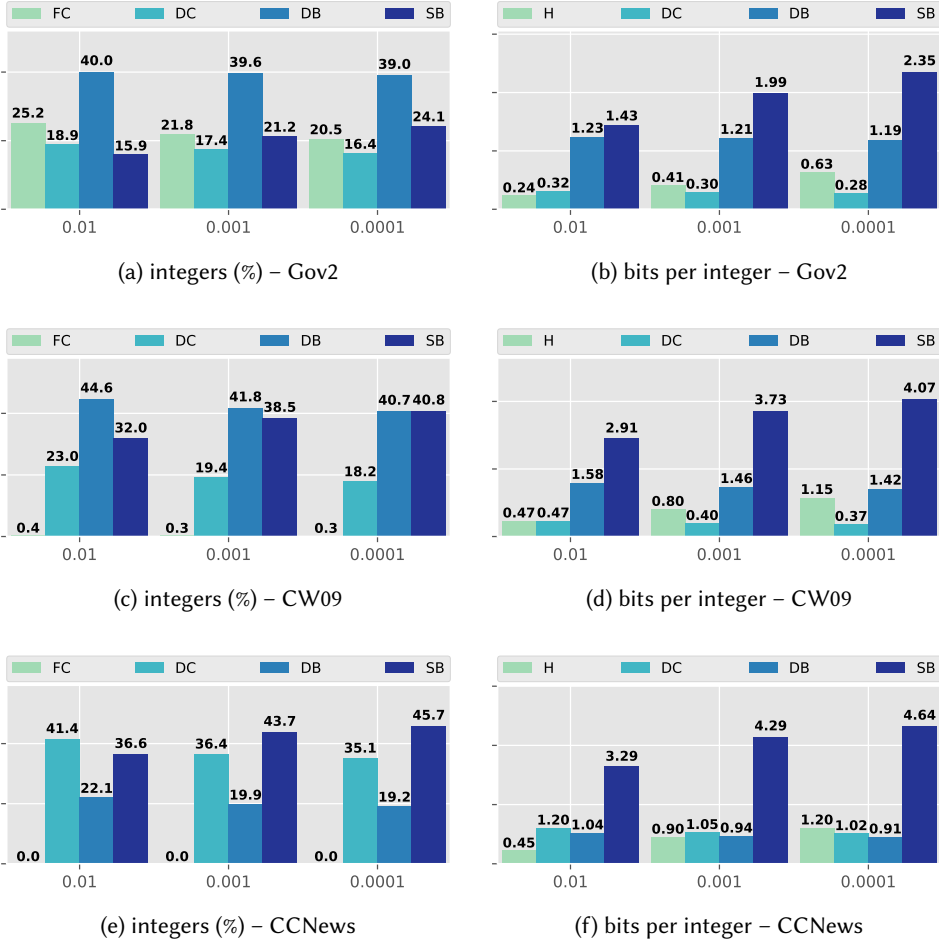

    \centering
    
    \subfloat[integers ($\%$) -- {\gov}]{
    \includegraphics[scale=0.6]{{{figures/integers_breakdowns.gov2}}}
    }
    \subfloat[bits per integer -- {\gov}]{
    \includegraphics[scale=0.6]{{{figures/bpi_breakdowns.gov2}}}
    }
    
    \subfloat[integers ($\%$) -- {\clue}]{
    \includegraphics[scale=0.6]{{{figures/integers_breakdowns.clueweb09}}}
    }
    \subfloat[bits per integer -- {\clue}]{
    \includegraphics[scale=0.6]{{{figures/bpi_breakdowns.clueweb09}}}
    }

    \subfloat[integers ($\%$) -- {\cc}]{
    \includegraphics[scale=0.6]{{{figures/integers_breakdowns.cc}}}
    }
    \subfloat[bits per integer -- {\cc}]{
    \includegraphics[scale=0.6]{{{figures/bpi_breakdowns.cc}}}
    } 
    
\caption{Plots (a), (c) and (e) show the percentage of integers
covered by the full chunks (FC), dense chunks (DC),
dense blocks (DB) and sparse blocks (SB) of the
Slicing approach.
Plots (b), (d), (f) show, instead,
how the bits per integer
rate of Slicing is fractioned among headers (H),
dense chunks, dense blocks
and sparse blocks.
For all plots, we show how the breakdowns change
by varying density.
\label{fig:breakdowns}}
\end{figure}

Lastly, the {\slice} solution stands in a middle position between these two classes, costing roughly
$0.9 \div 2$ bits per integer more than the most effective methods.
In Fig.~\ref{fig:breakdowns} we report the detailed
breakdown of how the integers of the test collections
are covered by the different universe slices and
how the bits per integer rate is fractioned
among them. Not surprisingly, most of the space is
spent in the representation of the sparse slices
of size $2^8$ that roughly cover (an average of) the
$20\%$, $37\%$ and $42\%$ of the integers of
{\gov}, {\clue} and {\cc} respectively.
Another meaningful thing to notice is that more than
$20\%$ of the integers of {\gov} are just covered
by runs of $2^{16}$ elements and, thus, represented
implicitly (dense chunks), whereas this does not happen
on the less clustered {\clue} and {\cc}.

Table~\ref{tab:decoding} reports the average nanoseconds
spent per decoded integer, measured by calling the operation \textsf{decode}
for each list in the index.
The methods {\vb}, {\roar}, {\roaropt} and {\slice} are the fastest.
However, {\vb} decodes a stream of $d$-gaps and we skipped the final
prefix-summing scan in this experiment, whereas
{\roar}, {\roaropt} and {\slice} directly decode the values 
without the need of further processing (thus, the results
compare more favourably for {\vb}).
There is no appreciable difference between the decoding
times of {\roar} and {\roaropt}.
The other bit-aligned methods {\uniform}, {\opt} and
{\interp} are much slower, with the latter being the
least efficient of all.
In particular, the \texttt{ds2i} library API does not expose
a \textsf{decode}
operation, thus we implemented it for Elias-Fano-based methods.
In such methods, a partition can be
represented using one among three different encodings according to
its characteristics, namely its relative universe of
representation and size.
These encodings include Elias-Fano, a bitmap and
an implicit representation whenever the relative universe of a partition
is equal to its size (see~\cite{2013:vigna} and~\cite{2014:ottaviano.venturini} for details).
Thus, efficient decoding of Elias-Fano codes basically reduces to
reading negated unary codes; bitmaps are decoded using the same
procedures as used in {\slice} (using the built-in \texttt{ctzll} function);
implicit partitions are decoded with inexpensive \texttt{for} loops.
The sequential decoding speed of Elias-Fano-based
methods is, anyway, two times less than the one of the
fastest methods.

The {\interp} mechanism does not feature specific
optimizations, except when decoding runs of consecutive
integers and is, on average, one order
of magnitude slower than the fastest methods.

\begin{table}
\centering
\scalebox{\mytablescale}{\begin{tabular}{c rrr l rrr l rrr}
\toprule

Method
& \multicolumn{3}{c}{$d = 10^{-2}$}
&
& \multicolumn{3}{c}{$d = 10^{-3}$}
&
& \multicolumn{3}{c}{$d = 10^{-4}$}
\\

\cmidrule(lr){1-1}
\cmidrule(lr){2-4}
\cmidrule(lr){6-8}
\cmidrule(lr){10-12}

& {\gov} & {\clue} & {\cc}
&
& {\gov} & {\clue} & {\cc}
&
& {\gov} & {\clue} & {\cc}
\\

\cmidrule(lr){2-2}
\cmidrule(lr){3-3}
\cmidrule(lr){4-4}
\cmidrule(lr){6-6}
\cmidrule(lr){7-7}
\cmidrule(lr){8-8}
\cmidrule(lr){10-10}
\cmidrule(lr){11-11}
\cmidrule(lr){12-12}

{\vb}
& 3648 & 6671 & 16954
&
& 710 & 1591 & 3732
&
& 40 & 214 & 523
\\

{\uniform}
& 4652 & 8356 & 22818
&
& 856 & 1700 & 4455
&
& 40 & 192 & 530
\\

{\interp}
& 12169 & 23608 & 58349
&
& 2649 & 6377 & 14765
&
& 160 & 905 & 2323
\\

{\opt}
& 4380 & 7920 & 21710
&
& 826 & 1640 & 4185
&
& 40 & 190 & 490
\\

{\roar}
& 377 & 598 & 1138
&
& 99 & 232 & 353
&
& 10 & 57 & 98
\\

{\roaropt}
& 503 & 962 & 1338
&
& 128 & 331 & 395
&
& 13 & 75 & 115
\\

{\slice}
& 507 & 1080 & 2370
&
& 135 & 378 & 820
&
& 11 & 60 & 159
\\

\bottomrule
\end{tabular}}
{\aftertabspace}
\caption{Average microseconds per AND query by varying density $d$.}
\label{tab:and_queries}
\end{table}

\subsection{List queries: boolean AND/OR}\label{sec:list_queries}
We now consider the two fundamental list-level
queries of intersections (boolean AND)
and unions (boolean OR).
Again, for all methods the result of the query
is materialized onto a pre-allocated output buffer
of 32-bit integers, thus we slightly modify the
\texttt{ds2i} code base to do so (rather than just
\emph{counting} matching integers).
To ensure a fair comparison,
we also slightly modify the pair-wise intersection and union functions of \texttt{CRoaring},
because these always output a new Roaring data structure resulting
from the operation, thus including (potentially expensive)
memory allocations during the process.
Thus, our modification avoids memory allocation
but the result is accumulated in the
pre-allocated output buffer mentioned above.

\return
Table~\ref{tab:and_queries} shows the result for
intersections.
The net result is that indexes partitioned by universe, {\roar}, {\roaropt} and {\slice}, are significantly more efficient
than those partitioned by cardinality, thanks to their
``simpler'' intersection algorithm using
substantially less instructions and branches.
As discussed, in this context simplicity means that,
being aligned to the same
relative universe, bitmap intersections can be carried out
by a sequence of inexpensive bitwise AND 64-bit operations;
sorted array intersections can be accelerated using
SIMD-based algorithms.
This results in \perf{6}{51} faster execution for
$d = 10^{-2}$; \perf{5}{42} for $d = 10^{-3}$;
\perf{4}{23} for $d = 10^{-4}$.

As a further evidence of this fact, we report in Table~\ref{tab:perf_counts}
some performance counts collected with the \texttt{perf} Linux utility,
when executing the queries on the {\gov} datasets.
We choose to report the counts for the {\pef} method because
it is the one generally performing better among the PC solutions.
From the numbers reported in the table we can see that
both {\roar} and {\slice} perform significantly less instructions
and branches, for example, $10\times$ and $8\times$ less instructions
for $d=10^{-2}$ and $d=10^{-3}$ respectively,
thus confirming our previous claim about the increase of performance.
The {\pef} method is also ``data hungry'' compared to {\roar} and {\slice}
as it is clear from the high number of L1 cache loads.
This is explained by the frequent switching of partitions
for higher density values.
Observe that {\pef} is actually exploiting the
data cache well (for example, only 228$\times 10^6$ misses
out of 102 $\times 10^9$ loads in L1 for $d=10^{-2}$),
however, the higher number of L1 references
imposes a significant penalty.
Also observe that {\slice}
is generally slower than {\roar}
because of the further slicing into smaller partitions,
inducing more branches that are not easily
predicted and thus partially eroding the
instruction throughput.
In fact, the (intentionally) simpler design of {\roar} is a lot more advantageous for SIMD instructions:
to confirm this, we recompiled the
\texttt{CRoaring} library by disabling explicit
SIMD optimizations and {\roar} scored the same as {\slice},
so vectorization does the difference.
However, notice how the difference in efficiency vanishes for lower density values because most of the
skipping happens at a coarser level.
Furthermore, also observe that the use of run containers
in {\roaropt} prevents some SIMD optimizations~\cite{lemire2018roaring}, thus reducing or even
annulling the performance
gap between {\roaropt} and {\slice}.

\begin{table}
\centering
\scalebox{0.9}{\begin{tabular}{l SSS l SSS l SSS}
\toprule

Quantity
& \multicolumn{3}{c}{$d = 10^{-2}$}
&
& \multicolumn{3}{c}{$d = 10^{-3}$}
&
& \multicolumn{3}{c}{$d = 10^{-4}$}
\\

\cmidrule(lr){1-1}
\cmidrule(lr){2-4}
\cmidrule(lr){6-8}
\cmidrule(lr){10-12}

& {\pef} & {\roar} & {\slice}
&
& {\pef} & {\roar} & {\slice}
&
& {\pef} & {\roar} & {\slice}
\\

\cmidrule(lr){2-2}
\cmidrule(lr){3-3}
\cmidrule(lr){4-4}
\cmidrule(lr){6-6}
\cmidrule(lr){7-7}
\cmidrule(lr){8-8}
\cmidrule(lr){10-10}
\cmidrule(lr){11-11}
\cmidrule(lr){12-12}

instructions ($\times 10^9$)
& 440.99 & 41.1 & 40.3
&
& 79.5 & 9.9 & 10.3
&
& 3.6 & 1.7 & 0.85
\\

instructions/cycle
& 2.07 & 2.11 & 1.62
&
& 1.98 & 1.77 & 1.53
&
& 1.75 & 1.30 & 1.28
\\

branches ($\times 10^9$)
& 69.7 & 4.7 & 5.5
&
& 11.7 & 1.5 & 1.5
&
& 0.5 & 0.35 & 0.14
\\

L1 loads ($\times 10^9$)
& 101.99 & 7.8 & 5.98
&
& 18.3 & 2.1 & 1.83
&
& 0.85 & 0.45 & 0.18
\\

L1 misses ($\times 10^6$)
& 228.4 & 421.6 & 282.4
&
& 51.1 & 100.1 & 62.9
&
& 6.7 & 16.9 & 6.8
\\

LL loads ($\times 10^6$)
& 15.7 & 78.2 & 24.4
&
& 5.45 & 19.8 & 7.8
&
& 1.49 & 4.1 & 1.85
\\

LL misses ($\times 10^6$)
& 11.2 & 35.2 & 14.1
&
& 4.5 & 12.5 & 5.6
&
& 1.43 & 3.1 & 1.54
\\

\bottomrule
\end{tabular}}
{\aftertabspace}
\caption{Performance counts collected with the \texttt{perf} Linux utility
when executing AND queries on the {\gov} dataset, for the {\pef}, {\roar}
and {\slice} methods, by varying density.
\label{tab:perf_counts}}
\end{table}

\begin{table}
\centering
\scalebox{0.9}{\begin{tabular}{l rrr l rrr l rrr}
\toprule

Method
& \multicolumn{3}{c}{$d = 10^{-2}$}
&
& \multicolumn{3}{c}{$d = 10^{-3}$}
&
& \multicolumn{3}{c}{$d = 10^{-4}$}
\\

\cmidrule(lr){1-1}
\cmidrule(lr){2-4}
\cmidrule(lr){6-8}
\cmidrule(lr){10-12}

& {\gov} & {\clue} & {\cc}
&
& {\gov} & {\clue} & {\cc}
&
& {\gov} & {\clue} & {\cc}
\\

\cmidrule(lr){2-2}
\cmidrule(lr){3-3}
\cmidrule(lr){4-4}
\cmidrule(lr){6-6}
\cmidrule(lr){7-7}
\cmidrule(lr){8-8}
\cmidrule(lr){10-10}
\cmidrule(lr){11-11}
\cmidrule(lr){12-12}

{\slice} \emph{with} SIMD
& 507 & 1080 & 2370
&
& 135 & 378 & 820
&
& 11 & 60 & 159
\\

{\slice} \emph{without} SIMD
& 816 & 1959 & 5190
&
& 213 & 558 & 1344
&
& 13 & 72 & 203
\\

\bottomrule
\end{tabular}}
{\aftertabspace}
\caption{The performance of the {\slice} method
when executing AND queries \emph{with} and \emph{without}
the use of SIMD instructions.
Clearly, the first row of the table
corresponds to the last row of Table~\ref{tab:and_queries}.
\label{tab:and_slicing}}
\end{table}

\return
In Table~\ref{tab:and_slicing} we also investigate the
impact of SIMD instructions for the intersection of small
sorted array discussed in Section~\ref{subsec:operations}.
The experiment highlights two important facts, one being
the consequence of the other:
(1) the vectorization of small arrays pays off, as the results
for AND are significantly better with SIMD instructions
(roughly $2\times$ better for sufficiently dense sequences);
(2) most of the running time is actually spent in intersecting
small arrays (not surprisingly, since bitmaps require
essentially bitwise instructions that are very cheap).
The latter fact explains why the SIMD optimization is so effective
and is consistent with the breakdowns reported in Fig.~\ref{fig:breakdowns}.
Lastly, the effect of vectorization clearly tends to diminish for
smaller sequences, being usually the ones with lower density
values, as we can see by comparing the values reported in
the columns corresponding to $d=10^{-2}$ and $d=10^{-4}$.

\return
Table~\ref{tab:or_queries} shows instead the result for
unions. For the same reasons discussed above for
intersections, the indexes partitioned by universe
are superior.
However, due to the scan-based nature of unions, the
performance gap with respect to the indexes partitioned
by cardinality is not as high as the one for intersections.
It is anyway consistent and equal to
\perf{4.5}{13} for $d = 10^{-2}$;
\perf{3.7}{11.5} for $d = 10^{-3}$;
\perf{3.6}{10.6} for $d = 10^{-4}$.
Finally, notice that the results for {\roar}, {\roaropt}
and {\slice} are very similar in this case, with {\roaropt}
being slightly less efficient.

\begin{table}
\centering
\scalebox{\mytablescale}{\begin{tabular}{c rrr l rrr l rrr}
\toprule

Method
& \multicolumn{3}{c}{$d = 10^{-2}$}
&
& \multicolumn{3}{c}{$d = 10^{-3}$}
&
& \multicolumn{3}{c}{$d = 10^{-4}$}
\\

\cmidrule(lr){1-1}
\cmidrule(lr){2-4}
\cmidrule(lr){6-8}
\cmidrule(lr){10-12}

& {\gov} & {\clue} & {\cc}
&
& {\gov} & {\clue} & {\cc}
&
& {\gov} & {\clue} & {\cc}
\\

\cmidrule(lr){2-2}
\cmidrule(lr){3-3}
\cmidrule(lr){4-4}
\cmidrule(lr){6-6}
\cmidrule(lr){7-7}
\cmidrule(lr){8-8}
\cmidrule(lr){10-10}
\cmidrule(lr){11-11}
\cmidrule(lr){12-12}

{\vb}
& 7754 & 12480 & 21000
&
& 2173 & 3924 & 6191
&
& 285 & 920 & 1407
\\

{\uniform}
& 9540 & 17952 & 29600
&
& 2704 & 5495 & 8589
&
& 366 & 1300 & 2000
\\

{\interp}
& 21115 & 39190 & 63972
&
& 6369 & 12898 & 20408
&
& 899 & 3185 & 5042
\\

{\opt}
& 8900 & 17000 & 28349
&
& 2560 & 5230 & 8300
&
& 350 & 1252 & 1887
\\

{\roar}
& 1737 & 3570 & 5001
&
& 562 & 1360 & 1762
&
& 80 & 356 & 543
\\

{\roaropt}
& 1950 & 4215 & 5180
&
& 638 & 1657 & 1812
&
& 86 & 408 & 551
\\

{\slice}
& 1955 & 4040 & 7440
&
& 590 & 1315 & 2265
&
& 73 & 276 & 476
\\

\bottomrule
\end{tabular}}
{\aftertabspace}
\caption{Average microseconds per OR query by varying density $d$.}
\label{tab:or_queries}
\end{table}

\subsection{Point queries: access and nextGEQ}\label{sec:point_queries}
For the methods {\vb}, {\uniform} and
{\interp}, the {\access}($i$) operation returns the
integer in position $i$ mod $B$ from the
partition of index $\lfloor i / B \rfloor$, for $B = 128$ integers in this experimentation.
In particular, the {\vb} method requires decoding
the partition and perform a prefix-summing
scan up to position $i$ mod $B$.
The {\opt} method needs to first locate the partition from which to return the integer because partitions have variable sizes.
Similarly, all solutions partitioned by universe,
{\roar}, {\roaropt} and {\slice}, have to take into account
the cardinality of each chunk covering a universe of 
size $2^{16}$ in order to locate the chunk containing the $i$-th integer.
Table~\ref{tab:access_queries} shows the timings
of such algorithms.

The {\uniform} method provides generally the fastest query time thanks to the constant-time random access algorithm of
Elias-Fano, with {\opt} and {\slice}
being in close second position.
The decoding operation performed by {\vb} imposes a
performance penalty with respect to such methods,
that is more evident for, clearly, sparser datasets.
Again, notice that the {\access} time decreases
for decreasing values of density, because fewer
partitions per encoded sequence are represented.
Lastly, {\slice} is faster than {\roar} because the latter adopts a linear search for the proper
chunk to access, whereas {\slice} builds cumulative cardinality counts.
Concerning the {\roaropt} variant with run containers,
the linear-search approach employed absorbs roughly 90\% of the time resulting in a significant slowdown, confirming
the experimental conclusions already given by the authors
of Roaring~\cite{lemire2018roaring}.

\return
Table~\ref{tab:next_geq_queries} shows instead the
results for the {\Nextgeq}($x$) query.
In this case, for all methods partitioned by
cardinality, the query is resolved by relying on
the skip pointers, as explained in Section~\ref{subsec:card}. 
Precisely, the wanted partition is first identified by binary
searching $x$ among the skip pointers, then the operation
is concluded in the partition.
Differently, the mechanism partitioned by universe
\emph{directly} identifies the partition
by considering fields of the binary representation
of the key $x$.
For this reason and as already discussed in Section~\ref{subsec:operations},
this operation is actually faster than {\access} for
universe-aligned methods.

Again, the Elias-Fano-based methods provides
generally better efficiency but with {\roar} and
{\slice} being faster especially for lower
density values: in such cases, {\slice}
is the fastest thanks to the further skipping introduced
within a single partition.
The slowdown imposed by the runs in {\roaropt} is alleviated by the use of binary search in this case.

\begin{table}
\centering
\scalebox{\mytablescale}{\begin{tabular}{c rrr l rrr l rrr}
\toprule

Method
& \multicolumn{3}{c}{$d = 10^{-2}$}
&
& \multicolumn{3}{c}{$d = 10^{-3}$}
&
& \multicolumn{3}{c}{$d = 10^{-4}$}
\\

\cmidrule(lr){1-1}
\cmidrule(lr){2-4}
\cmidrule(lr){6-8}
\cmidrule(lr){10-12}

& {\gov} & {\clue} & {\cc}
&
& {\gov} & {\clue} & {\cc}
&
& {\gov} & {\clue} & {\cc}
\\

\cmidrule(lr){2-2}
\cmidrule(lr){3-3}
\cmidrule(lr){4-4}
\cmidrule(lr){6-6}
\cmidrule(lr){7-7}
\cmidrule(lr){8-8}
\cmidrule(lr){10-10}
\cmidrule(lr){11-11}
\cmidrule(lr){12-12}

{\vb}
& 195 & 174 & 240
&
& 155 & 184 & 222
&
& 105 & 151 & 189
\\

{\uniform}
& 118 & 122 & 173
&
& 88 & 103 & 123
&
& 58 & 75 & 86
\\

{\interp}
& 890 & 835 & 1295
&
& 904 & 960 & 1230
&
& 685 & 876 & 1062
\\

{\opt}
& 154 & 171 & 210
&
& 118 & 145 & 126
&
& 77 & 100 & 72
\\

{\roar}
& 475 & 545 & 610
&
& 294 & 453 & 402
&
& 111 & 365 & 310
\\

{\roaropt}
& 5604 & 18710 & 2852
&
& 2151 & 7681 & 1221
&
& 443 & 2254 & 612
\\

{\slice}
& 153 & 170 & 244
&
& 105 & 116 & 152
&
& 55 & 61 & 78
\\

\bottomrule
\end{tabular}}
{\aftertabspace}
\caption{Average nanoseconds per random access query by varying density $d$.}
\label{tab:access_queries}
\end{table}

\begin{table}
\centering
\scalebox{\mytablescale}{\begin{tabular}{c rrr l rrr l rrr}
\toprule

Method
& \multicolumn{3}{c}{$d = 10^{-2}$}
&
& \multicolumn{3}{c}{$d = 10^{-3}$}
&
& \multicolumn{3}{c}{$d = 10^{-4}$}
\\

\cmidrule(lr){1-1}
\cmidrule(lr){2-4}
\cmidrule(lr){6-8}
\cmidrule(lr){10-12}

& {\gov} & {\clue} & {\cc}
&
& {\gov} & {\clue} & {\cc}
&
& {\gov} & {\clue} & {\cc}
\\

\cmidrule(lr){2-2}
\cmidrule(lr){3-3}
\cmidrule(lr){4-4}
\cmidrule(lr){6-6}
\cmidrule(lr){7-7}
\cmidrule(lr){8-8}
\cmidrule(lr){10-10}
\cmidrule(lr){11-11}
\cmidrule(lr){12-12}

{\vb}
& 252 & 226 & 308
&
& 255 & 226 & 279
&
& 197 & 181 & 243
\\

{\uniform}
& 187 & 122 & 250
&
& 146 & 155 & 175
&
& 91 & 113 & 120
\\

{\interp}
& 955 & 897 & 1385
&
& 951 & 1012 & 1290
&
& 710 & 878 & 1100
\\

{\opt}
& 167 & 182 & 229
&
& 138 & 157 & 144
&
& 94 & 118 & 89
\\

{\roar}
& 115 & 137 & 185
&
& 90 & 119 & 133
&
& 55 & 80 & 82
\\

{\roaropt}
& 105 & 138 & 188
&
& 80 & 115 & 136
&
& 50 & 72 & 85
\\

{\slice}
& 145 & 174 & 225
&
& 90 & 110 & 134
&
& 48 & 57 & 69
\\

\bottomrule
\end{tabular}}
{\aftertabspace}
\caption{Average nanoseconds per nextGEQ query by varying density $d$.}
\label{tab:next_geq_queries}
\end{table}

\section{Conclusions}\label{sec:conclusions}

The problem of introducing a compression format
for sorted integer sequences, with good practical 
intersection/union performance, is old but pervasive
in Computer Science, given
its many applications,
such as web search engines to mention a notable one.
%
Identifying a single solution to the problem is not generally
easy, rather the many space/time trade-offs
available can satisfy different application requirements
and the ``best'' solution should always be determined
by considering the actual data distribution.
To this end, we compare the two different paradigms that 
partition an inverted list for efficient query processing, 
either \emph{by cardinality} or \emph{by universe}.

\begin{figure}
\centering
\includegraphics[scale=0.75]{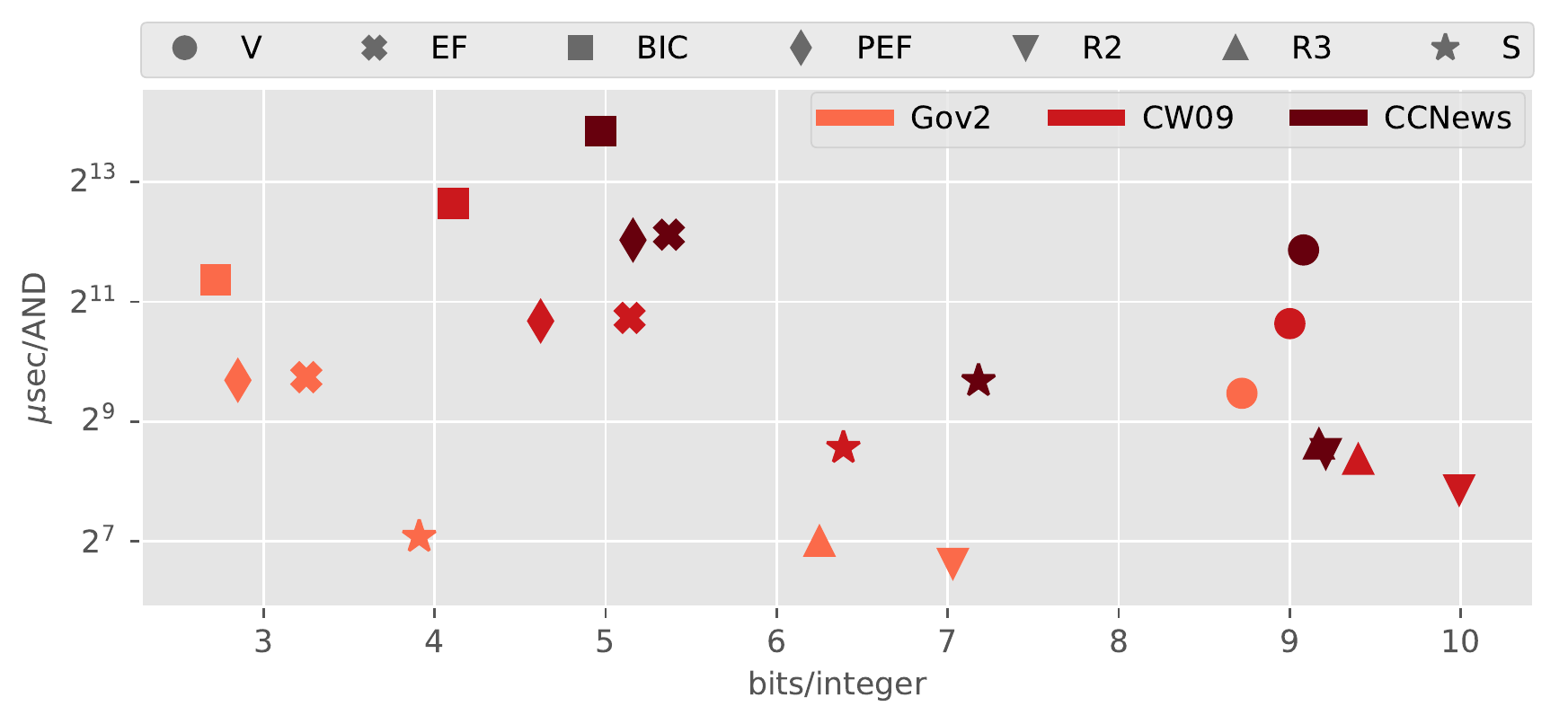}
\caption{Space/time trade-off curve for the different
representations summarized in Table~\ref{tab:strategies},
when executing AND queries under a density level of $10^{-3}$.
Note the logarithmic scale on the $y$ axis.
\label{fig:pareto}}
\end{figure}

Figure~\ref{fig:pareto} is a clear summary of such
experimental comparison because it shows different
space/time trade-off
points achievable for the list intersection operation which 
is the core one for inverted indexes.
On the one hand, techniques that use a partitioning-by-cardinality approach offer the best space effectiveness,
such as Elias-Fano-based methods and Interpolative;
on the other hand, the partitioning-by-universe
paradigm offers a remarkably improved intersection
efficiency at the expense of space effectiveness, as apparent with the Roaring method.
The Slicing solution devised here offers a leading compromise
between these two edge points,
by combining operational efficiency
with space effectiveness.
Observe that the Variable-Byte mechanism is generally 
dominated by other space/time trade-off points: its main
strength lies in the simplicity of the implementation
and the remarkably compact corresponding code 
(as far as SIMD
instructions are not considered).

Because of the maturity reached by the state-of-the-art
and the specificity of the problem,
identifying future research directions is not immediate.
We mention some promising ones.

In general, devising ``simpler'' compression
formats that can be decoded with algorithms using
low-latency instructions (e.g., bitwise) and with as few
branches as possible, is a profitable line of research,
as demonstrated by the experimentation in this article.
Such algorithms favour the super-scalar execution of modern CPUs and are also suitable for SIMD instructions.

Another direction could look at devising
\emph{dynamic and compressed} representations for integer
sequences, able of also supporting additions and deletions.
This problem is actually a specific case of the more general \emph{dictionary problem}, which is a fundamental
textbook problem.
While a theoretical solution already exists
with all operations supported in optimal time
under succinct space~\cite{PV17CPM},
an implementation with good practical performance could
be of great interest for dynamic inverted indexes.

\begin{acks}
This work was partially supported by the BIGDATAGRAPES project
(grant agreement \#780751, European Union's Horizon 2020 research and innovation programme).
\end{acks}

\renewcommand{\bibsep}{2.0pt}
\bibliographystyle{ACM-Reference-Format}
\bibliography{bibliography} 

\end{document}